# Charge Density Wave Phase Transition on the Surface of Electrostatically Doped Multilayer Graphene

**Authors:** *Gen Long[†], Shuigang Xu[†], Ting Zhang[†], Zefei Wu, Wing Ki Wong, Tianyi Han, Jiangxiazi Lin, Yuan Cai, Ning Wang\**

**Affiliations:**

*Department of physics and the Center for 1D/2D Quantum Materials, The Hong Kong University of Science and Technology, Clear Water Bay, Hong Kong, China*

\*Correspondence to: phwang@ust.hk

[†] These authors contributed equally to this work.

**We demonstrate that charge density wave (CDW) phase transition occurs on the surface of electronically doped multilayer graphene when the Fermi level approaches the M points (also known as van Hove singularities where the density of states diverge) in the Brillouin zone of graphene band structure. The occurrence of such CDW phase transitions are supported by both the electrical transport measurement and optical measurements in electrostatically doped multilayer graphene. The CDW transition is accompanied with the sudden change of graphene channel resistance at $T_m = 100K$, as well as the splitting of Raman G peak ($1580$ cm$^{-1}$). The splitting of Raman G peak indicats the lifting of in-plane optical phonon branch degeneracy and the non-degenerate phonon branches are correlated to the lattice reconstructions of graphene -- the CDW phase transition.**

According to the band structure of graphene, the density of state (DOS) diverges at the M points in the first Brillouin zone (BZ). These points are known as van Hove Singularities (VHSs) [1]. When the Fermi energy $E_F$ is approaching to the VHSs, the DOS divergence may lead to system instabilities and phase transitions [2, 3]. Low-energy states such as charge density waves (CDW) may occur. The VHS features in Ca-intercalated few-layer graphene [4, 5], artificially twisted bilayer graphene [2] and high-quality single layer graphene under a high magnetic field [6] have been studied theoretically and/or experimentally. However, the possibility of CDW phase transitions that involve VHSs in electrostatically doped graphene has been a subject of considerable debate since the discovery of graphene because extremely large doping concentrations are difficult to achieve to tune the Fermi level of graphene to VHSs.

To effectively tune the Fermi level of graphene, we fabricate multilayer graphene field-effect double-layer transistors (EDLTs) biased through ionic liquid gating. A high density of charge carriers are effectively introduced into graphene via electric double layers formed at the interface between the ionic liquid and the topmost graphene surface [7]. Once the Fermi level of graphene is tuned to the VHSs by the ionic liquid gating, a sudden increase of the graphene channel resistance is repeatedly detected at about 100K, similar to the features of CDW phase transitions [8]. Evidently, the splitting of the Raman G peak further demonstrates the lattice reconstructions, which is associated with the CDW phase transitions when the sample temperature is below the critical point ($T_m$=100K) while the Fermi energy is close to VHSs [8, 9].

**Graphene electric double-layer transistors (EDLTs)**

EDLT devices are chosen to achieve the CDW phase transition through effective tuning of the Fermi energy of graphene to the VHSs electrostatically (Supplementary Materials). Hall devices with side gate electrodes are fabricated using standard electron beam lithography (see the inset in Fig.1(b)). A droplet of ionic liquid (DEME-TFSI) is applied onto the surface of graphene channel and the side gate electrodes (Section 1, Supplementary Materials). Under a positive (or negative) bias voltage at room temperature, cations (or anions) from the ionic liquid are derived onto the surface of graphene channel (Fig.1(a)). The inducing carrier efficiency (estimated based on the equivalent capacitance) is $\sim 5.23 \times 10^{13}$ cm$^{-2}$V$^{-1}$ by evaluating the Hall effect of the devices (Fig. 1(c)). The gating efficiency of the ionic liquid is approximately two orders of magnitude higher than that of widely used 300 nm-thick SiO$_2$ ($\sim 10^{11}$cm$^{-2}$V$^{-1}$). Importantly, the carrier density linearly depends on the gate voltage (Fig.1(c)) and does not depend on the thickness of graphene samples. Therefore, the gate voltage dependence of carrier density can be estimated expediently and accurately.

Figure 1(b) shows the transport characteristics measured from a typical EDLT at room temperature (the source-drain current $I_{ds}$=50nA) for forward/reverse gate sweeps. Carrier mobility in this device is approximately 2630 cm$^2$V$^{-1}$s$^{-1}$. The maximum resistance (near the charge neutrality point (CNP)) is approximately 12 *K*Ω with an on/off ratio >10. These properties demonstrate a reliable EDLT device. The differences in maximum resistance are attributed to the accumulation of charge impurities during measurement [10]. No chemical reaction occurs between the ionic liquid and graphene because the data collected are reproducible with a negligible leak current (~1nA). In addition, the resistance between drain and source remains unchanged after bias voltage is applied for over 5h (Fig.S1(c)). Stable EDLTs at high bias voltages (~4.5V) are critical for probing VHSs in graphene. We increase the gate voltage using small (~10mV) and steady steps (long waiting time of 2 seconds for each step) to minimize the disturbance to the ionic liquid from biasing voltages. This process allows the ionic liquid to remain stable at high biasing voltages. Similar to the dispersion of introduced charge carriers in other layered materials[11, 12], the introduced charge carriers are confined within the Thomas-Fermi (TF) screening length $l_{TF} = 2\pi/q_{TF}$, where $q_{TF} = \frac{2\pi e^2}{\varepsilon_r \varepsilon_0} DOS$ stands for the TF wave vector and $\varepsilon_r \varepsilon_0$ denotes the dielectric constant[13]. For quantitative estimation, we take DOS=0.1 (eV unit cell)$^{-1}$ which is the minimum value of the DOS around VHS supplied in ref.[14] (the maximum value of TF screening lengths should be obtained from the minimum DOS value). We obtain the TF screening length $l_{TF}$ = 0.0145 nm which is much smaller than the interlayer space of graphene (~0.35 nm). This means that most of the charge carriers induced by electric double layer are filled on the topmost graphene layer.

**CDW phase transitions near VHSs**

This high gating efficiency as well as reliabilities and stabilities of our devices enables the observation of VHSs in monolayer, bilayer and multilayer graphene (Section 3, Supplementary Materials). Tuning the Fermi level to VHSs enables us to verify the CDW phase transitions in heavily doped graphene. CDW transitions are usually probed through measurement of structure changes, conductivity variation, the optical characterization of Raman scattering [8, 9]. In the present work, the transport properties of graphene EDLTs with different thicknesses (0.34 nm/ monolayer~ 20 nm/ 59 layers) are studied at different temperatures. Multilayer graphene samples with thickness less than 10 nm (1~29 layers) prepared on $SiO_2$ substrates normally exhibit a metallic behavior at different temperatures. No CDW transition associated phenomenon is observed due to the influence of charge disorder on $SiO_2$ substrates as discussed later. Multilayer graphene with thicknesses greater than 14 nm consistently exhibit a sudden change in conductance at a specific temperature of approximately $T_m$=100K (Fig. 2(a)). To verify whether the sudden change in conductance is an intrinsic property of graphene, we have investigated the resistance variation of samples at different gate voltages (Fig. 2(b)). When the gate voltage is lower than 2 V (the Fermi level is much lower than VHSs), no sudden changes of resistance occurs. By increasing the gate voltage to 3.5 V (the Fermi level is close to VHS), we observe a sudden change of the channel resistance. These gate voltage (or carrier density) dependent features can be repeatedly observed in our sample (independent of the measurement sequence). These results eliminate the possibility that the sudden change in resistance is caused by the variation of the electrode contacts of the EDLT. Meanwhile, the ionic liquid should not cause any breakpoints because its melting point (the liquid-solid phase transition) is 220 K[15], which is much higher than $T_m$. Therefore, the sudden changes in resistance are intrinsic properties of graphene resulted from a phase transition in the sample. All the gate voltages are applied at 250K (above the melting point of the ionic liquid) at which the ions in the liquid gate can move freely. The observed sudden change in resistance is consistent with the typical feature of the CDW phase transitions [8]. A CDW transition has been detected by measuring the resistivity of 1T-$TaS_2$[16].

To investigate the influence of the charge disorder effect originating from the $SiO_2$ substrate on the graphene CDW transition, we prepare EDLTs with graphene thicknesses smaller than 10 nm directly on h-BN flakes. It is well known that the rough surfaces of $SiO_2$ substrates cause strong charge impurities and thus degrade the quality of graphene devices [17]. We find that thick h-BN sheets with layer numbers larger than 11 (~7nm) are sufficient for screening the charge impurity effect from $SiO_2$ substrates [18]. However, a too thick h-BN causes difficulties in fabricating EDLT devices. We choose a 3.5 nm-thick multilayer graphene prepared on an 8 nm-thick h-BN. A clear sudden change of the channel resistance is observed at 102K (Fig. 2(a)). The CDW phase transitions occurring at the topmost layer are fragile and very sensitive to the charge impurities and surface roughness from the substrates. Graphene layers below the topmost layer (along with h-BN sheets) can screen these disorder effects effectively.

The CDW phase transitions in multilayer graphene are further reflected by Raman spectroscopy. The Raman spectrum obtained from the 16 nm-thick sample (Fig.3(a)) exhibits a normal G peak at 130K (above $T_m$) or $V_{LG}$=0V. When $V_{LG}$ is increased to 3V and sample temperature is 90K (below $T_m$), the G peak splits into two peaks, which are denoted as $G^+$ and $G^-$, respectively. The G peak of Raman spectrum is related to two degenerate phonon branches: in-plane longitudinal optical branch (iLO) and in-plane transverse optical branch (iTO). The eigenvectors of these two phonon branches are shown in the inset of Fig. 3(a)[19]. The splitting of G peak implies that the degeneracy of iLO and iTO phonon branches is lifted, further supporting the model of lattice reconstruction in graphene. A similar behavior of Raman G peak splitting has been observed in graphene under uniaxial strains, in which lattice deformations are generated [20, 21]. As one of the direct consequence of CDW phase transitions, lattice reconstructions should be observed when cooling down graphene EDLTs [8]. The observed $T_m$=100K in our EDLTs fits in this temperature range of 90K to 130K. At $T_m$=100K the normal phase of graphene should have the same system energy with Graphene in the CDW phase. But at a lower temperature, graphene in the CDW phase is an energy-preferred structure. A systematic study of Raman shifts and angle-resolved normalized Raman intensity reveals that the splitting of G peak exhibits an isotropic nature (Fig.3(b)). The isotropic G peak splitting may indicate that the CDW phase transitions existing in multilayer graphene are in the form of domain structures with different crystallographic orientations that commonly exist in material phase transitions. Because the Raman spectroscopy detects the averaged signals of the irradiated graphene areas and the anisotropic features from each domain are smeared by each other.

The CDW phase transition can be ascribed to the Fermi surface nesting effect [3, 8, 22-25] when the Fermi surface is around VHSs (also called Van Hove nesting effect). The nesting vectors which connect the Fermi level states around M and M' points, will be the basis vectors of graphene reciprocal lattice in CDW phase (red solid arrow in Fig. 4). The nesting vector can be folded back to the origin of first BZ, i.e. the Γ point. In this case, the electron-phonon scattering can happen on the phonons around the Γ point which is consistent with the observed splitting of Raman G peak of graphene in CDW phase. The Raman G peak of graphene is associated with $E_{2g}$ mode at the Γ point in reciprocal space [19], and the lattice distortions arising from CDW phase transitions should result in variations of the G peak. Since iTO and iLO optic modes correspond to the vibrations of one sublattice against the other, the lifting of the degeneracy of iTO and iLO modes can be correlated to a dimerization pattern of the neighboring carbon atoms. Since electron-phonon coupling plays a critical role in CDW phase transitions, an enhancement of electron-phonon coupling may result in an increased critical temperature.

In summary, we have experimentally investigated the phase transitions on the surface of electrostatically doped multilayer graphene and demonstrated a phase transition which is consistent with CDW states. Multilayer graphene EDLT devices provide an effective way to electrically dope graphene and tune the Fermi level to VHSs. The CDW transition is accompanied with the sudden change of graphene channel resistance at Tm= 100K, and the Raman spectroscopy data provide additional evidence for the lattice reconstruction. The CDW phase transition is accounted for the Fermi surface nesting effect.

**Acknowledgement**

We acknowledge helpful discussions with Y. H. Chan, C. R. Pan and M. Y. Chou.

Financial support from the Research Grants Council of Hong Kong (Project Nos. 16302215, HKU9/CRF/13G, 604112 and N_HKUST613/12) and technical support of the Raith-HKUST Nanotechnology Laboratory for the electron-beam lithography facility at MCPF are hereby acknowledged.

**Author contributions:**

N. Wang and G. Long conceived the project. G. Long fabricated the devices and performed cryogenic measurements with the help of S. G. Xu. G. Long, T. Zhang and N. Wang analyzed the data and wrote the manuscript. Other authors provided technical assistance in the project.

**Competing financial interests:**

The authors declare no competing financial interests.

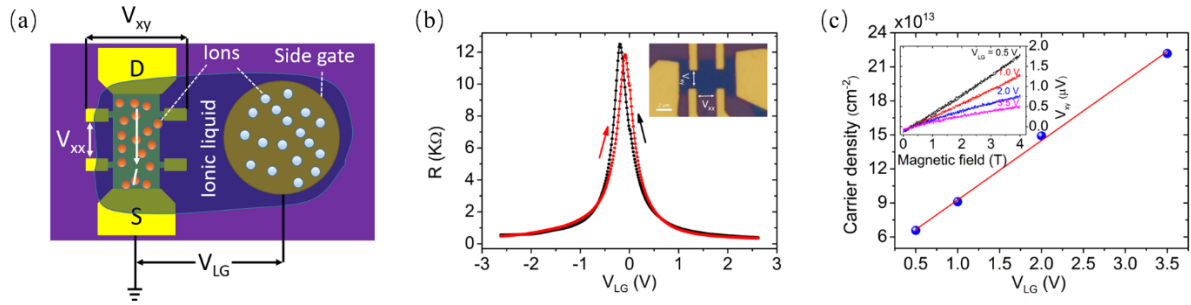

**Fig.1 Graphene EDLT and its ultra-high efficiency to induce charge carriers.** (**a**) EDLT device and measurement configuration. (**b**) Transport curves of EDLT at 260K (red: forward scanning; black: reverse scanning). The inset shows the optical micrograph of a typical graphene EDLT device. The scale bar is 2 $\mu m$. (**c**) The carrier density induced by varying gate voltages (blue dot) and the linear fitting result (red line). The inset shows the Hall voltage varies with magnetic fields at different gate voltages (black: 0.5V; red: 1.0V; blue: 2.0V; pink: 3.5V).

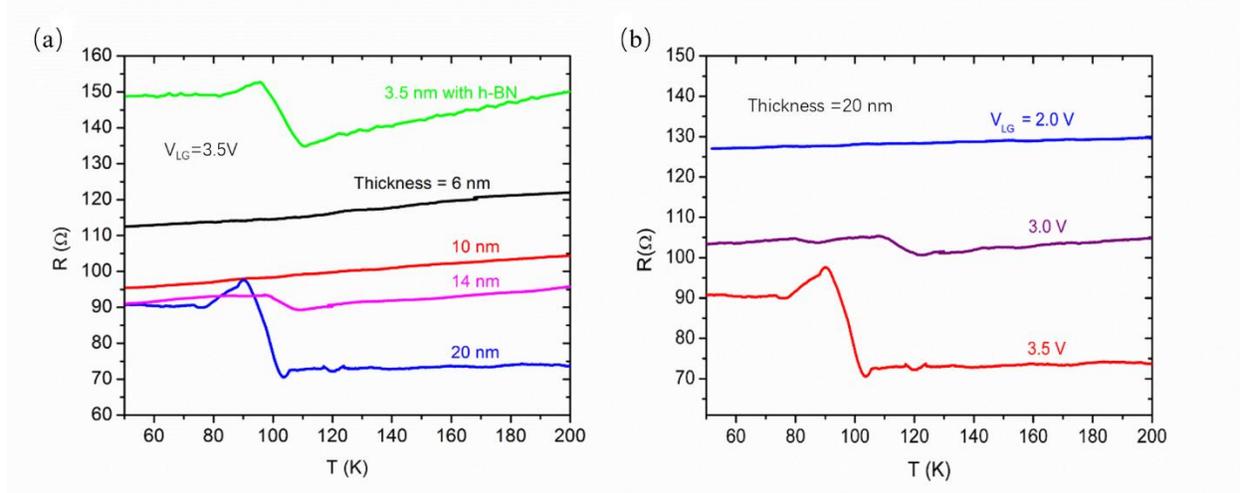

**Fig.2 Temperature dependence of EDLT channel resistance.** (**a**) Temperature dependence of channel resistance when the Fermi level is near VHS (gate voltage: 3.5V) with different thickness (green: 3.5nm; black: 6nm; red: 10nm; pink: 14nm; blue: 20nm). The green line represents the resistance of device prepared on BN substrate. (**b**) Temperature dependence of 20 nm-thick graphene channel resistance at different gate voltages (red: 3.5V; violet: 3.0V; blue: 2.0V).

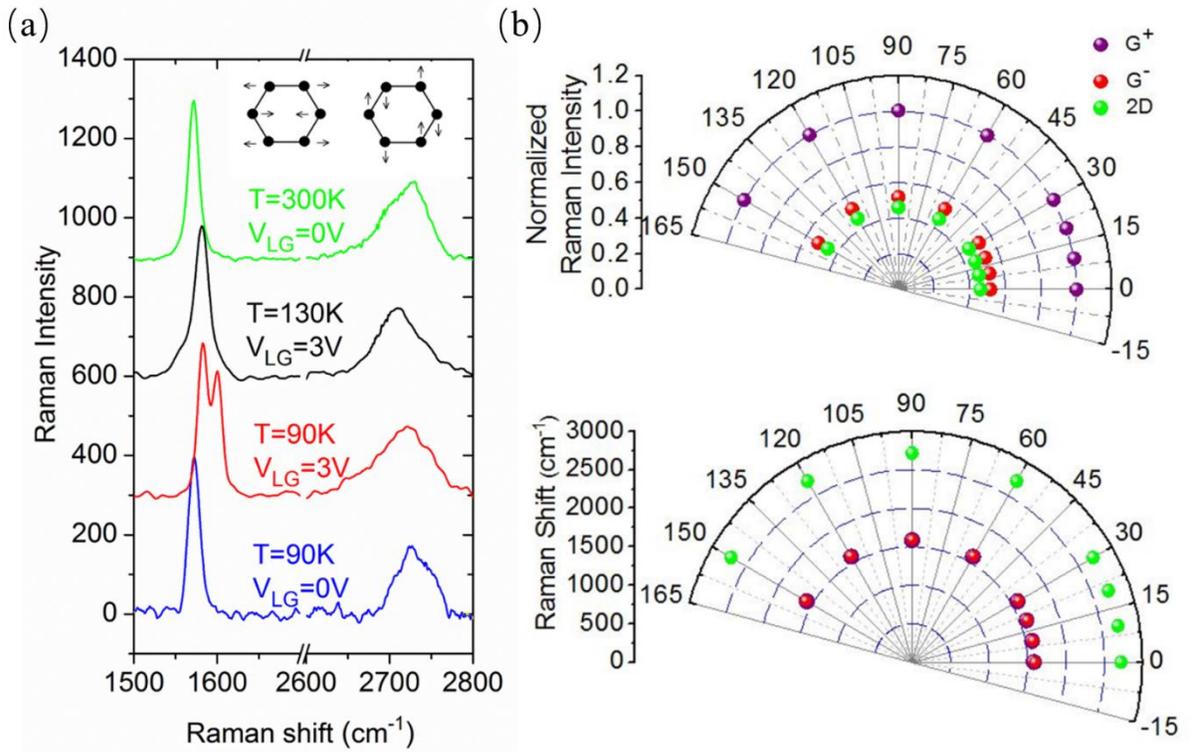

**Fig. 3 Raman spectroscopy of graphene EDLTs.** (**a**) The Raman spectrum of the 16 nm-thick sample under different situations (From top to bottom: T=300K, $V_{LG}$=0V; T=130K, $V_{LG}$=3V; T=90K, $V_{LG}$=3V; T=90K, $V_{LG}$=0V). (**b**) Angle-resolved normalized Raman intensity (Top) and Raman shifts (Bottom) of graphene in CDW phase. The intensity of Raman peaks are normalized to $G^+$.

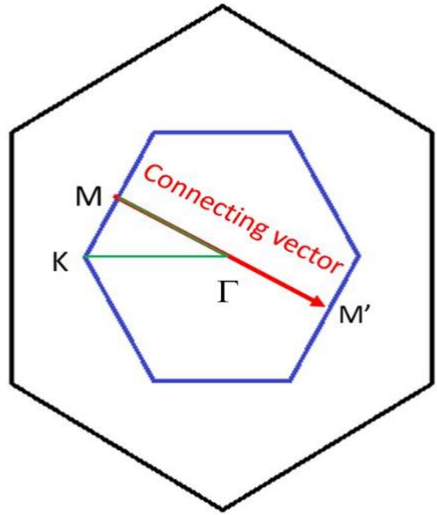

**Fig. 4 The two dimensional BZ of graphene.** As the Fermi level is tuned to M point VHS, the nesting vectors will be from M to M' (red arrow), and therefore in reduced BZ the nesting vector is at Γ point.

**Supplementary Materials for:**

# Charge Density Wave Phase transitions in Electrostatically Doped Multilayer Graphene

**Authors:** *G. Long, S. G. Xu, T. Zhang, Z. F. Wu, W. K. Wong, T. Y. Han, J. X. Z. Lin, Y. Cai, N. Wang\**

*Correspondence to: phwang@ust.hk

**Contents**

1. Device Fabrications
2. Raman measurements
3. Probing VHSs in Monolayer, Bilayer and Multilayer Graphene
4. Gate voltage and Fermi level

1. **Device Fabrications**

Thin flakes of graphene are prepared by micro-mechanical exfoliation of a single crystalline graphite. Graphene samples are first identified through the color contrast under an optical microscope. Then, the sample thickness is verified by Raman spectrometry (for few layer graphene) (Figs.S1A) and atomic force microscopy (sample thicker than 5 *nm*).

Standard electron-beam lithography, electron-beam evaporation and lift-off techniques are used to fabricate standard Hall devices with gate electrodes. The metal electrodes consist of Ti/Au/SiO$_2$ (5*nm*/80*nm*/30*nm*). All the six metal electrodes of Hal devices are covered by 30 nm-thick SiO$_2$ to isolate the electrodes from direct contact with the ionic liquid. The area of the side-gate electrodes are much larger (area ratio >$10^3$) than that of the samples. Ti is used to ensure good electrical contacts to graphene and Au aims to avoid oxidation of the electrodes. SiO$_2$ isolates the electrodes from any electrical contact with the ionic liquid (Figs.1A and B). A small constant AC current (50 *nA*) is applied between the drain and source electrodes to perform transport measurements. The transverse ($V_{XY}$) and longitudinal ($V_{XX}$) voltages along the Hall bar devices are measured simultaneously as a function of varied parameters (e.g., the gate voltage $V_{LG}$, temperature $T$ and or magnetic field $H$) using lock-in amplifiers. Because of the high density of ions at the interface between the ionic liquid and graphene samples as well as the thin electric layer (~1 *nm*) formed between the ionic liquid and graphene surface, an extremely strong electric field (~$5 \times 10^9$ *V/m*) is generated to introduce an extremely high density of charge carriers to the samples. A large area ratio (>$10^3$) between the side gate and the samples (Fig.S1B) is designed to make sure that the voltage drop is effectively applied at the interface between the sample and the ionic liquid.

We use the ionic liquid: N, N-diethy1-N-(2-methoxyethy1)-N-methyl ammonium bis-(trifluoromethylsulfony1)-imide (DEME-TFSI) for all the measurements of the EDLTs. Moisture has profound impacts on the performance of the ionic liquid, which destructs the reversibility and reliability of the devices. To reduce the moisture content in the ionic liquid, we have to heat the ionic liquid to 100 $^oC$ in a low pressure environment for 10 hours and all the measurements are performed under vacuum (~ $10^{-3}$ Torr) conditions. The ionic liquid is applied to the devices in a glove box. If moisture is induced during the fabrication process, a large leak current(much larger than $I_{ds}$) through the side gate electrode exists. In this case, we need to stop the measurement immediately. To confirm the reliability of samples, we measure the resistance consecutively while maintaining the bias voltage on over 5 hours. Fig.S1C shows that our samples are reliable and the possibility of chemical reaction between the ionic liquid and graphene has been ruled out.

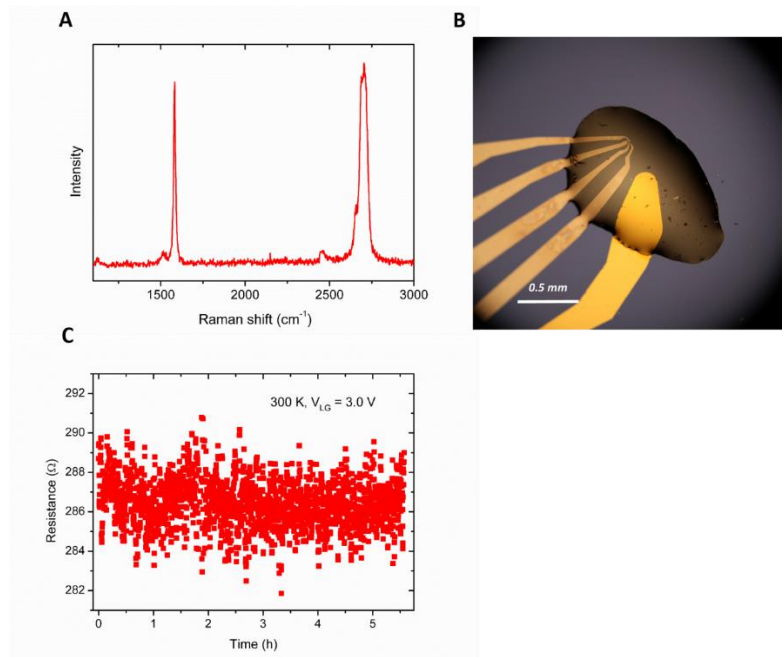

**Fig. S1.** (**A**) The Raman spectrum taken from bilayer graphene. (**B**) The optical micrograph of a typical graphene device. The length of scale bar is 0.5 mm. (**C**) One typical resistance profile varying with time while maintaining the gate voltage at 3 *V* at 300 *K*. The resistance is independent of time.

2. **Raman Measurements**

    Raman measurements are performed in a vacuum chamber ($10^{-7}$ torr) equipped with a quartz window. The sample temperature can be controlled by adjusting the heating power and the flow rate of liquid nitrogen (the freezing medium). A homemade holder is used to connect the side gate electrode and source electrode (or drain electrode) to the high and low output terminals of the gate voltage supplier respectively. This equipment allows us to measure the Raman spectra of our samples at different gate voltages and different temperatures. The D peak cannot be observed in our samples, indicating that the samples have minimum defects. The wavelength of the laser is 514 *nm* and laser power is 0.5 *mW*. Due to the low power of laser, the influence of the laser on sample temperature is ignored.

3. **Probing VHSs in Monolayer, Bilayer and Multilayer Graphene**

    Due to the large equivalent capacitance of electric double layers, the fine features in transport curves of graphene EDLT may be easily ignored if a large gate voltage step of 100 mV, for example, was used[1-4]. The small gate voltage steps used in this study allows the detection of every fine structure. A small step of the gate voltage is also important for probing VHSs because the full width at half maximum (FWHM) of the resistance changes at VHS is about 300 mV for graphene EDLTs.

    Figures S2A and S2B show the overall features of the measured resistance, which are consistent with previously reported results[5]. The insets in these two figures illustrate

resistance variation as a function of gate voltages at electron and hole sides, respectively. The appearance of troughs on electron and hole sides clearly denotes rapid variations of DOS when the gate voltage is set to around 3.8V and -3.8V, respectively. To better understand the occurrence of the troughs in transport curves of graphene EDLT, we calculate the derivative of conductance with respect to the gate voltages on both electron and hole sides. The conductivity of our samples is described through $\sigma = \mu n_{2D} e$, where $\mu$ denotes the carrier mobility, $n_{2D}$ is charge carrier density, and $e$ represents the elemental charge. The conductivity derivative can be written as $\frac{d\sigma}{dV_{LG}} = e \cdot \frac{d(\mu n_{2D})}{dV_{LG}}$, where $V_{LG}$ denotes the gate voltages. According to a previous report, the dependence of carrier mobility can be expressed as $\mu = \frac{8e}{h} \frac{1}{(n_{2D} \cdot n_s V_s^2 / \hbar^2 v_F^2) + 0.39 n_l}$, where $n_s$ and $V_s$ denote the density and scattering strength of short-range scattering impurities, respectively; $n_{2D}$ refers to the charge carrier density; $v_F$ denotes the Fermi velocity; and $n_l$ refers to the density of long-range scattering impurities[6]. When the gate voltage is high enough, the corresponding carrier density $n_{2D}$ becomes sufficiently large and dominates over $n_l$, so the derivative of mobility with respect to carrier density scales with square carrier density, $\frac{d\mu}{dV_{LG}} \sim \frac{1}{n_{2D}^2}$. We obtain $\frac{d\sigma}{dV_{LG}} = e\mu \cdot \frac{dn_{2D}}{dV_{LG}} + e n_{2D} \cdot \frac{d\mu}{dV_{LG}} \simeq e\mu \cdot \frac{dn_{2D}}{dV_{LG}}$ because the second term scales with $\frac{1}{n_{2D}}$.

Considering the definition of DOS, $n_{2D} = \int_0^{E_F} DOS\, dE$; hence, we obtain

$$\frac{d\sigma}{dV_{LG}} \simeq e\mu \cdot \frac{d\left(\int_0^{E_F} DOS dE\right)}{dV_{LG}} = e\mu \cdot DOS \cdot \frac{dE_F}{dV_{LG}} \quad \text{........................(S1).}$$

The absolute values of $\frac{d\sigma}{dV_{LG}}$ for electron (hole) doped graphene are shown in Fig.S2C (Fig.S2D). The linear dependence of $\frac{d\sigma}{dV_{LG}}$ on $V_{LG}$ near the CNP presented in the insets of Fig.S2C and D are consistent with that reported previously[7]. Since $\frac{dE_F}{dV_{LG}}$ decreases smoothly and monotonically with increasing gate voltages and has no contribution to peak-like features, the sudden increase in $d\sigma/dV_{LG}$ (marked by black arrows) in Figs.S2C and S2D directly indicate a rapid variation of DOS, which is the characteristic of VHSs in DOS. We define these peak centers (denoted by green arrows in Fig.S2C and S2D) as VHSs. The corresponding gate voltage deviating from the charge neutrality point on the electron (or hole) side is 3.66 V (or −3.98 V), with a high concentration of charge carriers (~1.91 × $10^{14}$ cm$^{-2}$ for electrons and 2.08 × $10^{14}$ cm$^{-2}$ for holes). The carrier density on the hole side is larger than that on the electron side because graphene exhibits band structure asymmetry[8]. Theoretical calculations show that the VHS on the hole side is farther away from CNP compared with that on the electron side.

Alternatively, the VHS can be detected through the Fermi level dependence of DOS. The dependence of Fermi energy on gate voltages can be derived roughly from the band structure of monolayer graphene (Section 4, Supplementary Materials). Translating gate voltages into Fermi energy, Eq. (S1) can be rewritten as

$$\frac{d\sigma}{dE_F} = e\mu \cdot \frac{d\left(\int_0^{E_F} DOS dE\right)}{dE_F} = e\mu \cdot DOS \quad \text{……………..………(S2).}$$

The absolute values of $\frac{d\sigma}{dE_F}$ are shown in Figs.S2E and F. Given that $e\mu$ is a constant, the peaks (denoted by green arrows) presented in Figs.S2E and F indicate the saddle-like features of DOS, that is, the VHS. Besides, the VHSs can be observed even in multilayer graphene EDLTs under a similar carrier density with that in monolayer graphene EDLT(Fig.S3).

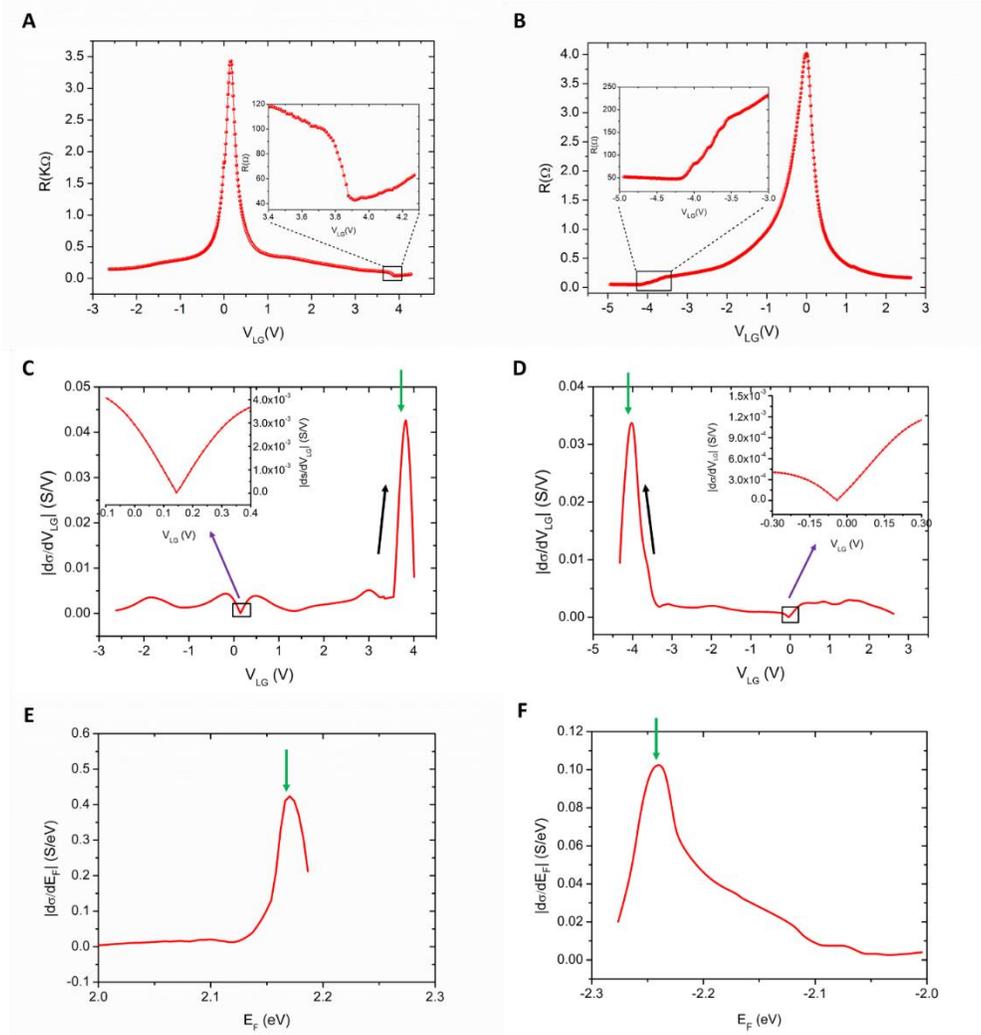

**Fig.S2 The VHS of graphene EDLT.** (A) and (B) Global features of EDLT transport curves. The black rectangles of (A) and (B) mark the position of troughs-like features on electron and hole sides, respectively. The insets show the zoomed in results of troughs-like features. (C) and

(D) The absolute values of the derivative of conductance with respect to $V_{LG}$ on hole and electron sides, respectively. The insets show the local features near CNP and the black arrows denote the sharp increase of curves. (E) and (F) show the absolute values of the derivative of conductance with respect to the estimated Fermi energy (E: electrode side; F: hole side). The green arrows in (C), (D), (E), (F) point out the positions of VHSs.

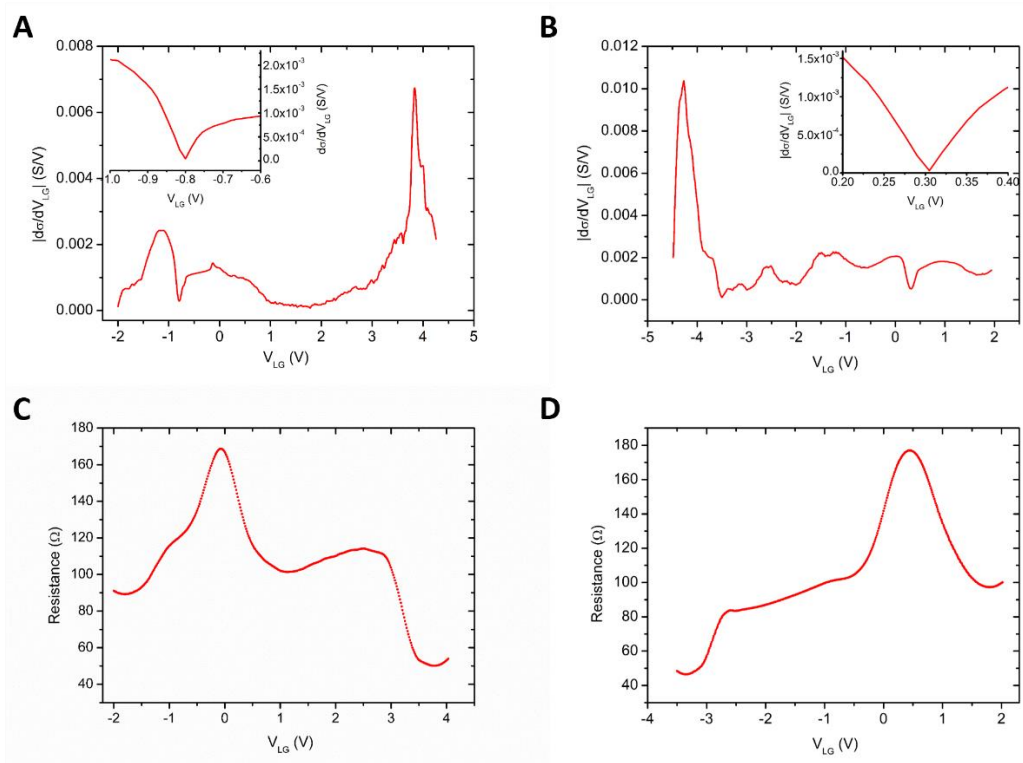

**Fig. S3. Van Hove Singularities of graphene with different thicknesses.** (A) and (B) show the absolute values of the derivatives of bilayer graphene's conductance with respect to $V_{LG}$, $|d\sigma/dV_{LG}|$ at hole and electron sides respectively. The insets show the local features near the neutrality points. (C) and (D) Transport curves of multilayer graphene.

## 4. Gate voltage and Fermi level

To estimate the Fermi level $E_{LG}$ dependence on gate voltage $V_{LG}$, the DOS of graphene has been calculated by both tight bonding approximation method (ignoring the next-nearest hopping) and DFT. For the sake of brevity, the tight bonding method is explained as an example here. By integrating DOS with Fermi energy, we get the Fermi energy $E_F$ dependence of carrier density $n_{2D} = \int_0^{E_F} DOSdE \equiv F_1(E_F)$. The gate voltage dependence of carrier density $n_{2D} = C_{eff}(V_{LG} - V_{th}) \equiv F_2(V_{LG})$ is obtained by measuring Hall Effect (Fig.1D). Now the carrier density works like a bridge between gate voltages and Fermi energy. The gate voltage dependence of Fermi energy can be written as $E_F = F_1^{-1} * F_2(V_{LG})$, where $F_1^{-1}$ stands for the inverse mapping of $F_1$.

Taking the distributions of local DOS as a uniform distribution, the obtained $E_F$ dependence on $V_{LG}$ is depicted by the red solid line of Fig.S4D. The obtained Fermi energy $E_F$ at VHSs are listed in table.S1 (Column titled as 'measured'), and the Fermi energy of VHSs obtained from calculations of band structures are also listed in table.S1 (Column titled as 'theory'). The measured Fermi energies at VHSs are consistently smaller than the calculated ones because disorder inevitably occurs, which causes the fluctuation of local DOS. The DOS at VHS decrease with increasing density of disorder[9]. A low DOS is favorable for increasing the efficiency of introducing charge carriers to tune the Fermi level. Therefore, in practice, the Fermi level should be higher than that calculated from measured carrier density (normally, calculations are based on the DOS of pristine graphene). Besides, the fluctuation of local DOS may induce a decrease in graphene quantum capacitance[10], and consequently, an increase in voltage drop across the capacitor with the same charge carrier density. Therefore, the chemical potential of graphene with local DOS fluctuation should be higher than that with uniform local DOS, provided that the charge carrier densities in both cases are equal. This occurrence indicates that the Fermi level obtained from carrier density is underestimated.

Figure S4A shows the treatment for the fluctuation of LDOS. The sample flake is divided into N small regions. The quantum capacitance of these small regions are denoted by $C_{Qi}$ ($i=1,2,...N$). An equivalent circuit is shown in Fig.S4B. The total capacitance $C_T$ of the device is the summation of all $N$ capacitance of $C_{EL0}$ and $C_{Qi}$, $C_{EL0} = C_{EL}/N$ denotes the capacitance of electric layer between the ionic liquid and graphene in small regions. According to Ref.[10] the quantum capacitance can be expressed by

$$C_Q = (C_T^{-1} - (NC_{EL0})^{-1})^{-1} = \frac{N\sum_{i=1}^{N} C_{Qi}/(C_{Qi}+C_{EL0})}{\sum_{i=1}^{N} 1/(C_{Qi}+C_{EL0})}$$ while the quantum capacitance for uniform

LDOS is $C_{QU} = \sum_{i=1}^{N} C_{Qi}$. Taking the distribution of LDOS as a lognormal form[10, 11], the $C_Q/C_{QU}$ ratio is calculated as a function of $C_{EL}/C_{QU}$ (Fig.S4C). If $C_{EL}/C_{QU}$ is large enough (>1), $C_Q/C_{QU}$ is close to 1 which means the fluctuation have little influence on the quantum capacitance of graphene; while for the case that $C_{EL}/C_{QU}$ is small (<0.5), $C_Q/C_{QU}$ decreases sharply, meaning that the quantum capacitance decreases dramatically under the influence of LDOS fluctuation. The electric layer capacitance can be considered as a constant (Fig.S4C insert) for varying gate voltages and excitation voltages. The $C_{QU}$ will increase with increasing gate voltages (increasing Fermi level of graphene). So, the fluctuation influence on quantum capacitance $C_Q$ will be enhanced at high gate voltages (high Fermi level). The solid blue line of Fig.S4D shows the Fermi energy dependence on gate voltages after considering the lognormal distribution of the LDOS. The corrected Fermi energies at VHSs are shown in table S1 (Column titled as 'corrected') and the corrected values are very close to the theoretical ones.

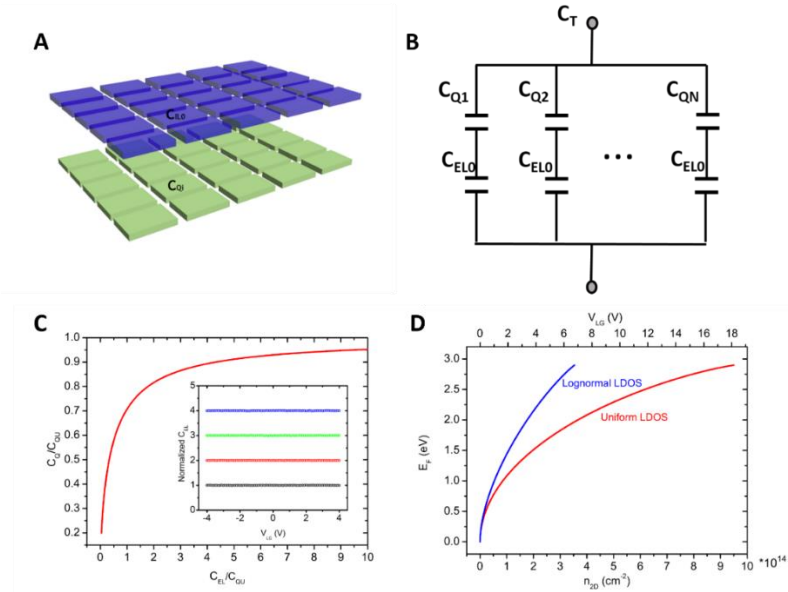

**Fig.S4. The gate voltage dependence of Fermi energy and the influence of LDOS fluctuation.** (**A**) Schematic diagram of LDOS. The EDLT is divided into N small areas. $C_{Qi}$ stands for the graphene local quantum capacitance of $i^{th}$ area they are different from each other due to the LDOS fluctuation. $C_{IL0}$ stands for the capacitance of electric layer of each small region. (**B**) An equivalent circuit of (**A**). $C_T$ stands for the total capacitance of the EDLT. (**C**) The $C_{EL}/C_{QU}$ dependence of $C_Q/C_{QU}$. The inset shows the normalized $C_{EL}$ varying with gate voltages with different excitation voltages (from top to bottom: 1000 mV, 100 mV, 10 mV and 1 mV, respectively). The curves have been offset for clarity. The normalized $C_{EL}$ equals to $C_{EL}$ divided by the average value of $C_{EL}$. The capacitance of the electric layer is independent of both gate voltages and excitation voltages. (**D**) The carrier density and gate voltage dependence of Fermi

energy. The red (blue) line is obtained without (with) considering the lognormal distribution of LDOS. At the same gate voltage, the Fermi energy for blue line is higher than the red line.

**Table S1.** The Fermi levels at the VHS obtained through TBA and DFT methods. The corrected results (after considering the fluctuation of LDOS) are presented.

|        | Electron side |              |           | Hole side    |              |           |
|--------|---------------|--------------|-----------|--------------|--------------|-----------|
| Method | Measured /eV  | Corrected /eV | Theory /eV | Measured /eV | Corrected /eV | Theory /eV |
| TBA    | 2.17          | 2.54         | 2.80      | -2.24        | -2.63        | -2.80     |
| DFT    | 1.28          | 1.52         | 1.63      | -1.40        | -1.67        | -2.31     |